\documentclass[10pt,twocolumn]{article}
\usepackage{cite,epsfig,verbatim,rotating,subfigure}
\usepackage{times}
\usepackage{fullpage,setspace}
\usepackage{url}
\usepackage[lined,boxed,commentsnumbered]{algorithm2e}
\usepackage{amsfonts}
\usepackage{pgfplots}

\begin{document}
\date{}
\pagestyle{empty}
\title{\Large\bf Mining Temporal Patterns from iTRAQ Mass Spectrometry(LC-MS/MS) Data}
\author{\begin{tabular}[t]{c@{\extracolsep{8em}}c}
Fahad Saeed$^*$\thanks{* Corresponding author: National Institutes of Health, 
Building 10, Room 6N312, 10 Center Drive, 
MSC 1603, Bethesda, MD 20892-1603.\mbox{\emph{ E-mail~address}: fahad.saeed@nih.gov}} & Trairak Pisitkun\\
Epithelial Systems Biology Laboratory & Epithelial Systems Biology Laboratory\\
National Institutes of Health (NIH) & National Institutes of Health (NIH) \\
Bethesda, Maryland 20892 & Bethesda, Maryland 20892 \\
\end{tabular}
\\
\\
\begin{tabular}[t]{c@{\extracolsep{8em}}c}
Mark A Knepper & Jason D Hoffert\\
Epithelial Systems Biology Laboratory & Epithelial Systems Biology Laboratory\\
National Institutes of Health (NIH) & National Institutes of Health (NIH) \\
Bethesda, Maryland 20892 & Bethesda, Maryland 20892 \\
\end{tabular}}

\maketitle

\begin{abstract}
\thispagestyle{empty} \singlespacing \textit{Large-scale proteomic analysis is emerging
as a powerful technique in biology and relies heavily on data acquired
by state-of-the-art mass spectrometers. As with any other
field in Systems Biology, computational tools are required to deal with this ocean of data.
iTRAQ (isobaric Tags for Relative and Absolute
quantification) is a technique that allows simultaneous
quantification of proteins from multiple samples. Although iTRAQ
data gives useful insights to the biologist, it is more complex to
perform analysis and draw biological conclusions because of its multi-plexed design.
One such problem
is to find proteins that behave in a similar way (i.e. change in abundance) among various
time points since the temporal variations in the proteomics data
reveal important biological information. Distance based methods
such as Euclidian distance or Pearson coefficient, and clustering
techniques such as k-mean etc, are not able to take into account
the temporal information of the series. In this paper, we present
an linear-time algorithm for clustering similar patterns among
various iTRAQ time course data irrespective of their absolute
values. The algorithm, referred to as Temporal Pattern Mining(TPM), maps
the data from a Cartesian plane to a
discrete binary plane. After the mapping a dynamic programming technique allows mining of similar data elements
that are temporally closer to each other. The proposed algorithm
accurately clusters iTRAQ data that are temporally closer to each
other with more than $99\%$ accuracy. Experimental results for
different problem sizes are analyzed in terms of quality of
clusters, execution time and scalability for large data sets. An
example from our proteomics data is provided at the end to
demonstrate the performance of the algorithm and its ability to
cluster temporal series irrespective of their distance from each
other. }
\end{abstract}
%\begin{keywords}
%Mass Spectrometery, iTRAQ labelling, Clustering, dynamic programming
%\end{keywords}

\singlespacing
\section{Introduction}
\label{sec:intro}

Mass spectrometry is a fundamental part of any modern proteomics research platform
for accurate protein identification and quantification \cite{hoffert2006}\cite{hoffert2}\cite{hoffert3}. Mass
spectrometers measure the mass-to-charge ratio(m/z) of ionized
particles \cite{mass1}. In the case of a typical LC-MS/MS proteomic experiment,
the ionized particles (i.e. peptides) are introduced into a mass spectrometer at the ion
source in the form of liquid solutions, then desolvated and
transferred into the gas phase as gas phase ions. A variety of search algorithms are
then used to match the peptide spectra to sequences in online
databases in order to identify the proteins in the mixture\cite{inspect}\cite{sequest}.

iTRAQ (isobaric tags for relative and absolute quantification) is a
technique used to identify and quantify proteins
from different sources in one single experiment. It uses isotope
coded covalent tags and is used to study quantitative changes in
the proteome\cite{itraq1,itraq2}. The method is based on covalent
labeling of the N-terminus and lysine side chains from
protein digestions with tags of various masses for distinction.
Up to 8 different tagging reagents (8-plex kit) are used to label
peptides from different samples. The samples are then pooled together as a single sample
and analyzed by mass spectrometer. The fragmentation of the attached tag
generates a low molecular mass reporter ion which is useful in
quantifying relative peptide abundance between the different iTRAQ channels.

The iTRAQ technique allows analysis of samples in a
more sophisticated and accurate manner in turn giving more
relevant biological information such as phosphorylation of
peptides or the effect of vasopressin at different time points in
the mass spec. Although the technique allows greater accuracy in quantitation, it raises many
computational problems. One such problem is to identify the
peptides that behave similarly for a given external agent e.g.
dDAVP over the time course study. Time course measurements from
iTRAQ data are becoming a common procedure in many systems
biology experiments
\cite{hierarcy-olsen}\cite{kmeans-forest}\cite{SOM-forest}.

If the experiment is subject to variations in time, the
conventional methods to cluster and analyze the similarity such as
Euclidean distance, Hamming distance have significant limitations.
Likewise the clustering mechanisms that use distance based
measures such as k-means\cite{kmeans}, or hierarchical
clustering\cite{hc} do not always succeed when responses are highly variable in magnitude. Other
methods such as fuzzy clustering of short time-series\cite{fuzzy}
are not computationally efficient after a certain number of time
courses due to the combinatorial explosion in possibilities. A facilitating characteristic
of successful scalable clustering is still to find a linear
algorithm that involves a small number of passes over the
database\cite{ghosh}.

In this paper we present a near-linear time clustering algorithm
that finds temporal patterns in a given large iTRAQ labeled
dataset using one or small number of passes over the data and
\emph{without compromising the quality of the clusters}. Our
algorithm draws its motivation from mapping problems in the parallel
processing community\cite{bokhari1}\cite{bokhari2} and
quantization in information theory\cite{shannon}\cite{ny}. The
proposed algorithm allows us to map the time points of a Cartesian
plane into a discrete plane. The mapping then produces a
predictable number of clustering possibilities that are then
binned using a efficient dynamic programming technique to extract
the patterns.

The paper is organized as follows. In section 2 we provide the
biological experimental details and the associated computational
problem. Section 3 discusses the proposed clustering algorithm and
complexity analysis. Section 4 discusses the experimental results
and illustrative examples for our clusters. Finally, conclusions
are presented in section 5 of the paper.

\section{Problem Statement}

\begin{figure}[htb]
\begin{center}
\vspace{-30mm}
\includegraphics[scale=0.50,angle=0]{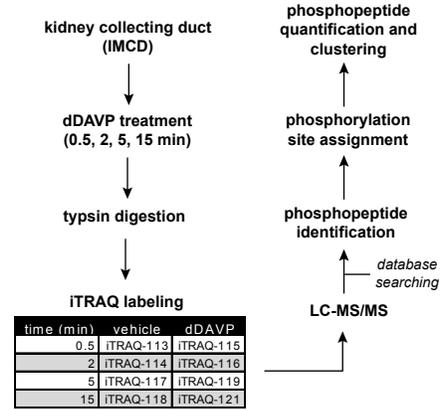}\vspace{-50mm}
\caption{\small \label{fig-exp} The Flow diagram for the
experiment}
\end{center}
\end{figure}

The objective of the study was to perform a quantitative
comparison of protein phosphorylation under vasopressin(dDAVP)
treatment using iTRAQ labels for different time points. LC-MS/MS
\cite{lc} phosphoproteomics analysis was performed and the time
course clusters that would be obtained from the study provides the
basis for modeling the signaling network involved. The flow
diagram for the experiment is shown in Fig.\ref{fig-exp}.

\begin{figure}[htb]
\begin{center}
\vspace{-10mm}
\includegraphics[scale=0.30,angle=0]{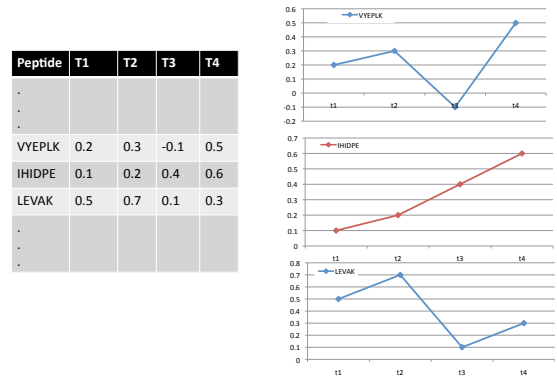}
\caption{\small \label{fig-star} The peptides with corresponding time scales and their
values. The first and third peptides belong to the same cluster}
\end{center}
\end{figure}

In computational terms, the problem that we wish to solve is as
follows. We are given a set of peptides with time points$(t_1, t_2
\cdots etc)$ with each time point having a certain real value
which in our case is the iTRAQ ratio. Given the peptides with time
points and real values, we want to be able to cluster the peptides
that give a similar pattern over the time course. An example of
such is shown in the figure \ref{fig-star}.

The first column in figure \ref{fig-star} shows peptide sequences followed by 4 columns of iTRAQ ratios corresponding to the
change in peptide abundance between the vasopressin and vehicle control
samples ("dDAVP/vehicle"), each corresponding to a different time point. Of course the number of columns would
be dependent on the time courses that are considered and can
increase or decrease depending on the particular experiment. Our
objective is to determine the data points that have similar
temporal patterns. It can be seen in the figure that peptide
$VYEPLK$ and $LEVAK$ have similar patterns over the time course
because both increase from point $t_1$ to $t_2$, decrease from
$t_2$ to $t_3$ and increase from $t_3$ to $t_4$. The peptide
$IHIDPE$ on the other hand increases for all the time points.
Hence, the peptides $VYEPLK$ and $LEVAK$ must be clustered
together.

Now let us formally define the problem. Let there be $N$ data points with each data point having
$K$ time course values and $X$ represents the peptide name.
Then, let $U$ present the set such that $U_i = \{X_i,(t_1,\cdots,t_K )\}$ where $1\leq i \leq N$.
Also, let the number of clusters be $Q$. Then, cluster set $C=\{c_1,c_2, \cdots, c_Q \}$ where each
$c_j \in U_i $ such that $1\leq i \leq N$ and $1\leq j \leq Q$. Then each of the cluster $c_j$
has the set of data points that have the same temporal pattern with time.

The temporal similarity is defined as follows. Let $c_{pq}$ represent $q^{th}$ data point in cluster
$p$ where $1 \leq p \leq Q$ and $1 \leq q \leq N$. Now let $c_{pq}(x)$ represent the \emph{mapped} data point at
point $x$. Let there be another \emph{mapped} data
point $c_{pq}(y)$ at point $y$ and $\forall x<y$. Now define an array
$r_h$ where $1 \leq h \leq K$. Each point in array

\begin{equation}
r_h=c_{pq}(y) - c_{pq}(x)
\end{equation}

Then, the data points that have strictly equal $r_h$ would be considered temporally similar.

\section{Proposed Mining Algorithm}

In this section, we present details of the proposed mining strategy, referred to as Temporal Pattern Mining (TPM). We also
analyze the computational complexities of the proposed algorithm.

The proposed algorithm \emph{TPM} draws its motivation from the mapping
problem in parallel and distributed computing \cite{bokhari1}\cite{khokhar1}\cite{bokhari2} and information
theory\cite{ny,shannon}. The mapping problem in parallel computing and mapping for our
mining algorithm share a similar characteristic. In mapping for
parallel processing, two sets of nodes are considered: problem
modules and processor modules. The objective is to map the problem
modules on to processor modules in an efficient manner. In mapping
for the mining algorithm we are seeking a mapping such that the
Cartesian plane of the data points are mapped onto a discrete
plane(e.g. binary plane) of finite possibilities. The discrete
plane is defined by using the Nyquist sampling technique that
allows conservation of the information from a continuous signal (or a
data set with real numbers). These finite possibilities, which can
grow exponentially with increasing time courses, are then mined
using our efficient dynamic programming technique. Algorithm\ref{algo1} gives an intuitive description of the
strategy.

\begin{algorithm}
\caption{Mapping based temporal pattern mining algorithm}\label{algo1}
%\begin{algorithmic}
\KwData{A set $U_i = \{X_i,(t_1,\cdots,t_K )\}$ of peptides and their time courses}
\KwResult{Compute the cluster set $C=\{c_1,c_2, \cdots, c_Q \}$  such that the clusters have temporal similarity within the distinct clusters}

\For{$i=1$ to $K$}{

Compute $A[i]:= mapping(U_i)$ from Cartesian to discrete plane

}

\While{there are values in $A$ that are not null}{

pick a random value from $A$ call it $A^R$

count++

\For{$j=0$ to $N$}{

            $distance = EditDistance(A^R, A[j])$

            \If {$distance == 0$}{
                $c_{count} \leftarrow A[j]$

\tcc{\emph{This is to eliminate values from A that have already been assigned to a cluster}}

                $A[j] \leftarrow NULL$
            }
        }
}
\end{algorithm}

\begin{algorithm}
\caption{Mapping function }\label{algo2}
\KwData{A data point in the Cartesian plane}
\KwResult{Return the discrete plane representation of the data point}

mapping(datapoint $U_w$)

Vector $V_w \leftarrow \emptyset$

\For{$w=0$ to $U_w.length()$}{

    \If{w=0}{

\tcc{\emph{save as name of the peptide}}

\Else{

        \If{$U_{w+1} - U_w \geq 0$}{

            $V_w = a$
        }
        \Else {$V_w = b$}

        }
}
}

\Return $V$

\end{algorithm}

\begin{algorithm}
\caption{Dynamic programming cluster extraction subroutine}
\label{algo3} \KwData{Two strings of length m and n}
\KwResult{LevenshteinDistance between the two string is returned }
\tcc{d is a table with m+1 rows and n+1 columns}

EditDistance(char s[1..m], char t[1..n])

   int $d[0..m, 0..n] \leftarrow \emptyset$

       \For{ i from 0 to m}{
         d[i, 0] := i
        \tcc{ deletion}
        }
           \For{j from 0 to n}{
        \tcc{insertion}
        d[0, j] := j
        }

   \For{ j from 1 to n}{

       \For{i from 1 to m}{

           \If{s[i] = t[j]}{
             d[i, j] := d[i-1, j-1]
        }

           \Else{

             d[i, j] := \emph{minimum}( d[i-1, j] + 1, d[i, j-1] + 1, d[i-1, j-1] + 1)

        }
        }

    }

  \Return d[m,n]

\end{algorithm}

\subsection{Mapping from Cartesian to discrete plane}
The proposed algorithm \emph{TPM} can be classified as a feature extraction
algorithm\cite{survey}. As defined in the problem statement in the
section above we have a number of peptides with
 associated iTRAQ ratio values and we are interested in
extracting clusters of the peptides that give similar expression levels(falling or rising) at different
time points. However, k-means or hierarchical clustering cannot be used because
the time points may be closer to each other in Euclidean distance, but may not be
close in temporal changes over the time course.

Clustering using the real values from the Cartesian plane however,
is not feasible because of its continuous nature (infinite
values). Therefore, with Cartesian co-ordinates the number of possible cluster
combinations will be infinite in nature and
clustering for all possible combinations is not computationally feasible. Thus, a
Cartesian plane co-ordinates has to be mapped to a more discrete
plane co-ordinates to restrict the number of combinations. The mapping function should be such that it
would allow us to quantify the variations in the data with respect
to time and also make the values discrete enough such that the number
of combinations that are possible would decrease drastically.

To address these challenges, the mapping function that is presented allows
us to make the values more discrete and also conserves the
important information of expression levels between the time
periods. Using the same notation that we presented in the previous
section. Let $U$ present the set of values such that $U_i =
\{X_i,(t_1,\cdots,t_K )\}$ where $1\leq i \leq N$. $X_i$
represents the peptides and $t_1, \cdots, t_K$ represents that
values of the expression level from $1$ to $K$. The mapping
from the Cartesian plane would be accomplished as follows:

\begin{equation}
[ M(x,y) = \left\{ \begin{array}{ll}
a & \mbox{if $(t_x < t_y)  and  (x<y)   and  ( y-x=1)$};\\
b & \mbox{o.w.}\end{array} \right. ]
\end{equation}

The assumption in the mapping function is that the first value is
zero. However, this is an assumption that is appropriate for our data but
is a not a generalized rule and can be changed accordingly. The
mapping function in its functionality is simple and does the
following. It looks for the data points at the next time point. If
the current data point is below the previous value it is assigned 'a'. Otherwise it is assigned a 'b'. Therefore, after the
mapping has been completed, each of the time series would be a
sequence of a's and b's with each of the characters representing
the rise or fall in the expression level. The number of
discrete levels can change according to the biological system
under consideration.

\begin{figure}[htb]
\begin{center}
\includegraphics[scale=0.30,angle=0]{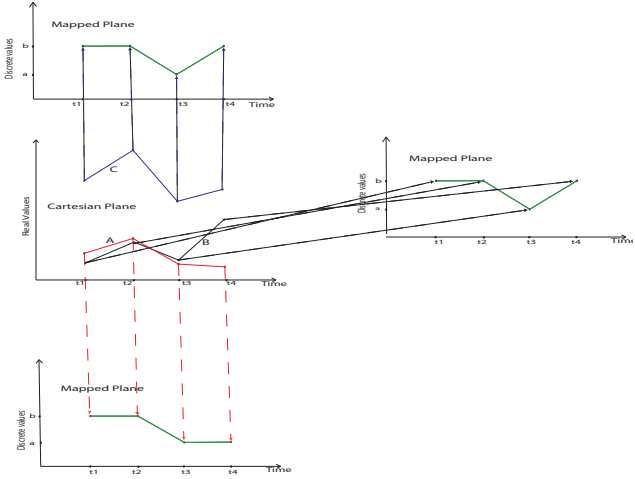}
\caption{\small \label{fig-mapping} The mapping of a Cartesian plane to a discrete mapped plane}
\end{center}
\end{figure}

The mapping function has been defined in a way that makes
a clear distinction between the data points that have similar
temporal patterns and the ones that don't. Figure \ref{fig-mapping} shows that
data points A(red) and B(black) are very close to each other in distance.
However, in our mining scenario we are more interested in the
pattern of the expression levels over time and in consideration of
our criterion, the data point C( blue) is closer to B(black). If a
naive k-means clustering would be executed, data points A and B
would be clustered together. Using our mapping function however,
we are able to make a clear distinction irrespective of the
Euclidean distance of the data points with respect to each other.
Observe that the mapping of co-ordinates that is produced by our
algorithm is same for the data points that have similar
temporal behaviors (i.e. B and C). Therefore, once the mapping has
been accomplished, the data points that have the same mapping
belong to a single cluster. The psuedocode for the mapping function is given in Algorithm \ref{algo2}.

\subsection{Efficient Extraction of Clusters}
Once the mapping is complete, the next step is to mine the
clusters that are present in the data set. An efficient way is
needed to extract the clusters from the mapped data because the
number of clusters can increase exponentially with increase in the
time points as well as the number of discrete levels desired.
Consider, for example, two discrete levels are considered as in
our case.  The number of clusters would be
upper bounded by the order of $Z^N$ (for $Z$ states or $2^N$ for
two states), but may or may not be present in the data\cite{kot}.
Even for a moderately large $N$ the number of combinations are huge and for each data
point going through all of the possible combinations is waste of
precious computing resources. For \emph{TPM} we developed an
efficient technique that allows us to keep the search space confined to the clusters that are present in the
dataset. Therefore, using our technique the exponential search
space of possible clusters is minimized to the set of clusters
that are present in the data set, saving valuable computing resources.

The procedure to mine the relevant clusters shown in Algorithm 3.
The crux of the technique is based on a dynamic programming edit
distance algorithm that allow us to calculate the 'distance' of a
particular data point from another. We randomly pick a data point
from the mapped data values. The randomly mapped data point is
then used to calculate the levenstein distance with other data
points. The data values that have zero distance with one another belong
to the same cluster, because they would have the same
pattern. The technique is very efficient in practice because for
large number of time points, the number of clusters that are
actually present in the data is far less than the possible number
of clusters. Thus, the computational complexity is greatly
decreased and makes the system more efficient.

\begin{figure}[htb]
\begin{center}
\includegraphics[scale=0.30,angle=0]{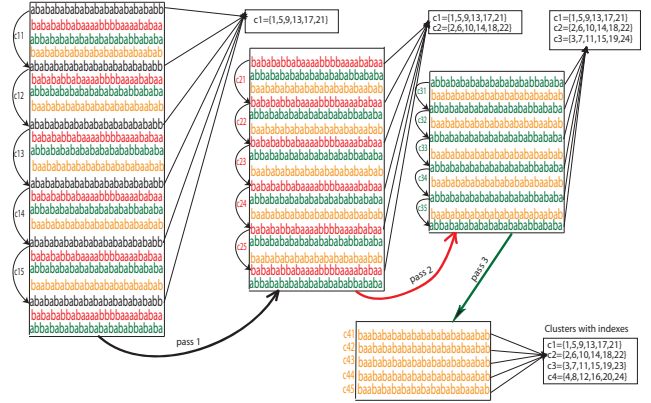}
\caption{\small \label{fig-tree} Extraction of clusters after mapping is complete}
\end{center}
\end{figure}

Figure \ref{fig-tree} shows the clusters are extracted from the
set in only small finite operations as compared to large number of
operations that would be required if all the possible combinations
would be considered. The figure shows $27$ time points for each
data set that has been mapped using our mapping strategy. The
total number of possible combinations are $2^{27}$ for each of the
data points making total operations equal to $23 \times 2^{27}$
i.e. $O(N2^N)$. However, considering our strategy, we pick one of
the data points randomly. In the figure, data points (time courses) in black font are picked and
the edit distance is calculated for each of the data points. These data
points would get a zero distance and would
be accumulated in a single cluster $c_1$. After, adding them to
the cluster set these black data points are excluded from the search
and the process is repeated until only one kind of data points is
present. The total number of operations is equal to $23 \times 4$
i.e. $O(QN)$ where $Q$ is the number of clusters and $N$ is the
number of elements present. It is very apparent in fig
\ref{fig-tree} that only 3 passes are needed to extract all the
clusters from the data set.

%\begin{figure}[htb]
%\begin{center}
%\includegraphics[scale=0.20,angle=0]{3dtree}
%\caption{\small \label{fig-tree3d} Extraction of clusters for higher dimension states}
%\end{center}
%\end{figure}

\small{The extraction of the clusters works very well in practice for a two-state dataset. However, it can be further improved by using
kd-tree data structure for extraction of clusters for higher
dimensional states\cite{kd1}\cite{kd2}\cite{kd3}. The k-d tree is
a multidimensional binary search mechanism that represents a
recursive subdivision of the data space into disjoint subspaces by
means of d-dimensional hyper planes. The root of the tree then
represents all the patterns, while the children of the root
represents subsets of patterns in the subspaces. Searching for the
clusters for the algorithm can then be performed in
$O(QN^{(1-1/No.of states)})$. Note that as the number of states
increase, the searching time would reduce correspondingly because
of the division of subspaces using the hyper planes. 
}

%Fig.
%\ref{fig-tree3d} shows a 3-dimensional kd-tree and gives an
%intuitive picture of how the subspaces are reduced for searching.
%The first split cuts the root cell into two sub cells, each of
%which is then split in to two sub cells. Further, each of these is
%split into two sub cells. In this way, the subspaces are reduced
%very dramatically decreasing the search times for higher
%dimensional states.

\subsection{Complexity Analysis}
The time complexity of the algorithm can be broken down into two
parts. The first part is for the mapping and the second part for
the extraction of the clusters. The mapping part has complexity of
$O(KN)$ since there are $N$ data elements and each is of length
$K$. The second part of the algorithm is for the extraction of the
clusters. This part run $Q$ times where $Q$ is the number of
clusters present in the data. For each run, dynamic programming
algorithm is executed, which runs in $O(K^2)$ time; assumption is
that both of the data points that are being compared are equal in
length. This procedure runs for $Q$ times giving the complexity of
$O(QK^2)$. Thus the total time complexity of the algorithm is
$T(.) = O(QK^2)+O(KN)$.

\section{Performance Evaluation}
Performance evaluation was done for the quality of the clusters
that were extracted as well as the efficiency of the technique. We
tested the algorithm with data from a large-scale quantitative
phosphoproteomics experiments done as follows: Inner medullary
collecting duct (IMCD) samples were incubated in the presence or
absence of 1nM dDAVP (vasopressin) for 0.5, 2, 5, and 15 minutes
(N=3) followed by LC-MS/MS-based phosphoproteomic analysis.
Quantification used 8-plex iTRAQ and commercially available
software. These phosphopeptides were analyzed with our algorithm
in order to identify groups that changed in abundance with similar
temporal responses after exposure to vasopressin. The algorithm
identified 16 clusters of phosphopeptides with distinct temporal
profiles. These time-course clusters provide a starting point for
modeling of the signaling network involved. The
algorithm\footnote{An executable can be
obtained by requesting the author; A webservice will also be available at 
authors' page.}has been implemented in Java(TM)
SE Runtime Environment $(build 1.6.0)$. The experiments were
conducted on a Dell server consisting of 2 Intel Xeon(R)
Processors, each running 2.40 GHz, with 12000 KB cache and 64GB
DRAM memory. The operating system on the server is Linux RedHat
enterprise version with kernel 2.6.9-89.ELlargesmip.

For the timing experiments we used the same data that we got from
our biological experiments. In order to access the timing for the
algorithm, we generated the data as follows. The complexity
analysis suggested that the algorithm must exhibit a linear time
with increasing data points. Therefore, we wanted to access the
timing by keeping other variables relatively constant i.e. the
number of clusters(Q) and the length of the time points(K). We
generated our data set for timing evaluation by replicating the
data that we got from our biological experiment. The ratio of the
number of clusters and the length of the time points remained
constant whenever we replicated our data by a real positive whole
factor. We tested the timings for data points from $3500$ to
$80,000$ elements.

\begin{figure}[htb]
\begin{center}

\begin{tikzpicture}
\begin{axis}[normalsize,legend pos= south east,xlabel=\textsc{Number of datapoints(N)},
ylabel=Time in seconds
 ]

\addplot+[black,sharp plot] coordinates
{(0,0) (3500,0.878)(7000,1.266) (14000,1.579) (20000,1.99) (40000,2.538) (80000,2.141)};

\addplot+[black,sharp plot] coordinates
{(0,0) (3500,0.866)(7000,1.34)(14000,1.78)(20000,2.300) (40000,2.732)(80000,2.737)};

\addplot+[black,sharp plot] coordinates
{(0,0) (3500,0.922) (7000,1.272) (14000,1.589) (20000,2.101) (40000,2.940) (80000,2.581)};

\addplot+[black,sharp plot] coordinates
{(0,0) (3500,0.563) (7000,1.372) (14000,1.789) (20000,1.76) (40000,2.440) (80000,3.281)};

\addplot+[black,sharp plot] coordinates
{(0,0) (3500,0.666)(7000,1.24)(14000,1.38)(20000,2.700) (40000,2.632)(80000,3.737)};

\legend{$Q=5$,$Q=10$,$Q=15$, $Q=20$, $Q=30$}

\end{axis}
\end{tikzpicture}

\caption{\small \label{fig-elements} Timing with increasing number of data points}

\end{center}
\end{figure}
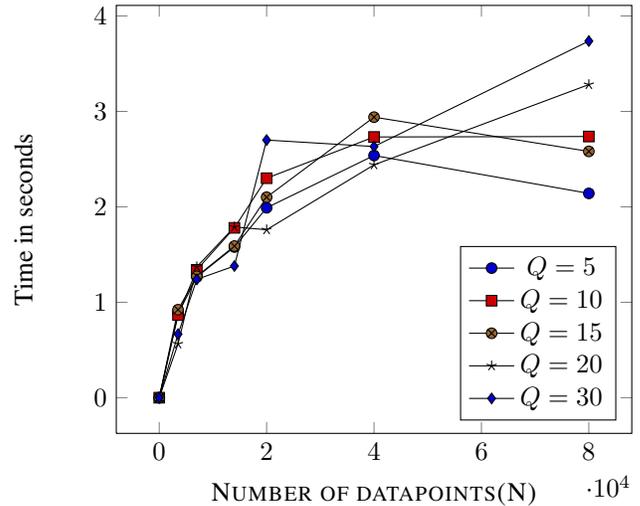

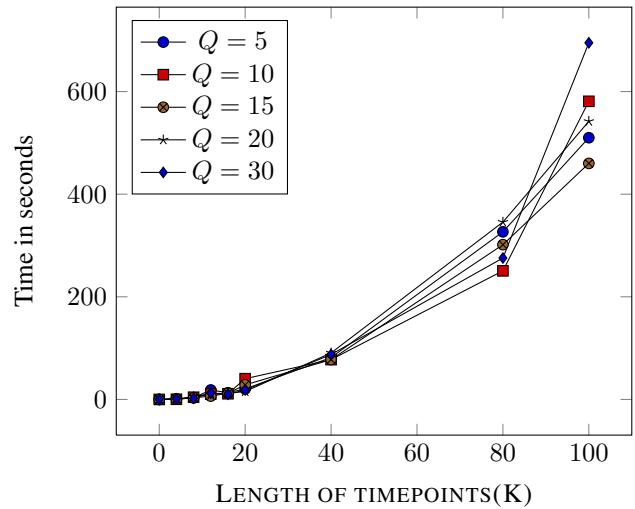
\begin{figure}[htb]
\begin{center}

\begin{tikzpicture}
\begin{axis}[legend pos=north west, xlabel=\textsc{Length of timepoints(K)},
ylabel=Time in seconds
 ]
\addplot+[black,sharp plot] coordinates
{(0,0) (4,0.826) (8,3.264) (12,18.466) (16,13.056) (20,20.4) (40,81.6) (80,326.6) (100,510)};

\addplot+[black,sharp plot] coordinates
{(0,0) (4,0.326) (8,4.264) (12,10.466) (16,11.056) (20,40.4) (40,77.6) (80,250.6) (100,581)};

\addplot+[black,sharp plot] coordinates
{(0,0) (4,0.726) (8,3.264) (12,6.466) (16,12.056) (20,28.4) (40,77.6) (80,301.6) (100,460)};

\addplot+[black,sharp plot] coordinates
{(0,0) (4,0.826) (8,5.264) (12,11.466) (16,12.056) (20,15.4) (40,90.6) (80,345.6) (100,542)};

\addplot+[black,sharp plot] coordinates
{(0,0) (4,1.126) (8,2.264) (12,12.466) (16,10.056) (20,18.4) (40,87.6) (80,275.6) (100,695)};

\legend{$Q=5$,$Q=10$,$Q=15$,$Q=20$,$Q=30$}

\end{axis}
\end{tikzpicture}

\caption{\small \label{fig-timespoints} Timing with increasing number of time points}

\end{center}
\end{figure}

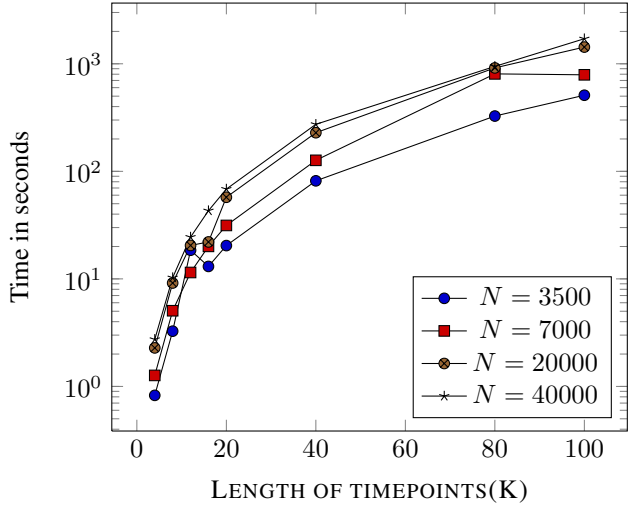
\begin{figure}[htb]
\begin{center}

\begin{tikzpicture}

\begin{semilogyaxis}[legend pos=south east, xlabel=\textsc{Length of timepoints(K)},
ylabel=Time in seconds
 ]
\addplot+[black,sharp plot] coordinates
{(0,0) (4,0.826) (8,3.264) (12,18.466) (16,13.056) (20,20.4) (40,81.6) (80,326.6) (100,510)};

\addplot+[black,sharp plot] coordinates
{(0,0) (4,1.266) (8,5.056) (12,11.466) (16,20.056) (20,31.4) (40,126.6) (80,805.6) (100,790)};

\addplot+[black,sharp plot] coordinates
{(0,0) (4,2.281) (8,9.152) (12,20.466) (16,22.056) (20,57.4) (40,228.6) (80,915.6) (100,1430)};

\addplot+[black,sharp plot] coordinates
{(0,0) (4,2.738) (8,10.264) (12,24.466) (16,43.056) (20,68.4) (40,273.6) (80,945.6) (100,1710)};

\legend{$N=3500$,$N=7000$,$N=20000$, $N=40000$}

\end{semilogyaxis}
\end{tikzpicture}

\caption{\small \label{fig-timespoints1} Timing with increasing number of time points}

\end{center}
\end{figure}

The timings for the algorithm with increasing number of data
points is shown in Fig. \ref{fig-elements} for up to $80000$. The
timings observed have a linear trend with increasing number of
data points as predicted by our complexity analysis. Even for up to
$80000$ elements the timings observed are no more than 3 seconds.
Observe, as the number of clusters decrease, the times observed
for the same number of data points have a decreasing trend. The
reason is that for fewer clusters, the number of passes that have
to be made to extract cluster pattern is smaller further
decreasing the computation time. It is also useful to know the
behavior of the algorithm with increasing number of time points(K)
in the data. Fig. \ref{fig-timespoints} shows the timings of the
algorithm with increasing number of time points(K) for up to 100
for variable number of clusters. The timings observed are in
accordance with our complexity analysis that suggested a $O(K^2)$
behavior and time remains under 500 seconds for clustering data
sets that have 100 time points. Figure \ref{fig-timespoints1} also
shows the timings with increasing number of time points while
varying the number of data points. As can be seen the algorithm
exhibits a very consistent behavior with increasing number of time
points as well as increasing number of data points. The constants
used during these experiments are $K=4, Q=30$ and $N=3500$
wherever appropriate.

\begin{table}
\caption{\small \label{fig-table} Number of correct clusters with
different algorithms}
\begin{tabular}{ ||l || c || r || }
\hline
  Algorithm & Clusters Identified & Correct Clusters \\
\hline
\hline
  K-means & 16 & 3 \\
\hline
Hierarchical & 18 & 6 \\
\hline
SOM & 20 & 10 \\
\hline
TPM & 16 & 16 \\
\hline
\end{tabular}
\end{table}

The quality of the clustering was then compared with other algorithms
such as kmeans\cite{kmeans-forest}\cite{kmeans}, Self
organizing maps\cite{SOM-forest}\cite{som1}, and Hierarchical
clustering\cite{hierarcy-olsen}\cite{h1}\cite{h2}. A brief
summary of the results from these experiments are in table
\ref{fig-table}. The assessment of the quality was done as follows. The standard
algorithms used as described in
\cite{kmeans-forest}\cite{SOM-forest}\cite{hierarcy-olsen} on our
dataset which are the same iTRAQ labeled data that has been
described in the literature. For the clusters that were identified
by these algorithms, if there were more than $1\%$ of data points
that were incorrect it is labeled as incorrect cluster. Although
there are many data points in the cluster that are similar to one
another; data points that are incorrectly clustered can have a
serious impact on the quantification of the proteins. As shown in
the table, the number of clusters that were identified varied
according to the algorithm used because of the high variability in
the data. For the cluster that were identified, above mentioned
criteria was used to distinguish the correct clusters from the
incorrect one. The results obtained matched well with the
previous studies e.g. In \cite{SOM-forest} the authors using self organizing maps(SOM) were only
able to use $9$ clusters that were sufficiently accurate to be
used for quantification studies thus limiting the biological
meaningful data analysis. Using our algorithm, it can be seen that
the number of clusters observed were in agreement with our theoretical analysis. All
of the data points that were included in the clusters didn't had a
variation more than $1 \%$ making it a highly accurate clustering
algorithm for iTRAQ labeled protein quantification analysis.

\begin{figure}[htb]
\begin{center}

\begin{tikzpicture}

\begin{axis}[legend pos=south east, xlabel=\textsc{Time in minutes},
ylabel=iTRAQ ratios
 ]

\addplot+[black,sharp plot] coordinates
{(0.5,-0.5326) (2,-2.0514) (5,-0.04268) (15,-0.37668)};

\addplot+[black,sharp plot] coordinates
{(0.5,-0.08789) (2,-.09482) (5,-0.01012) (15,-0.07871)};

\addplot+[black,sharp plot] coordinates
{(0.5,-1.00637) (2,-1.09876) (5,-.88202) (15,-0.95902)};

\addplot+[black,sharp plot] coordinates
{(0.5,-0.75222) (2,-0.76174) (5,-.419) (15,-.92117)};

\addplot+[black,sharp plot] coordinates
{(0.5,-.32438) (2,-.56875) (5,-.17155) (15,-.363)};

\addplot+[black,sharp plot] coordinates
{(0.5,-.22211) (2,-.37442) (5,-.02845) (15,-.22395)};

\addplot+[black,sharp plot] coordinates
{(0.5,-.03399) (2,-.2468) (5,-.03719) (15,-.10916)};

\addplot+[black,sharp plot] coordinates
{(0.5,-.44535) (2,-.5671) (5,-.05281) (15,-.2439)};

\addplot+[black,sharp plot] coordinates
{(0.5,-.04474) (2,-.31379) (5,-.00438) (15,-.5845)};

\addplot+[black,sharp plot] coordinates
{(0.5,-.13191) (2,-.31379) (5,-.00438) (15,-.05845)};

\end{axis}
\end{tikzpicture}

\caption{\small \label{fig-cluster1} Example of clustered data
points}

\end{center}
\end{figure}
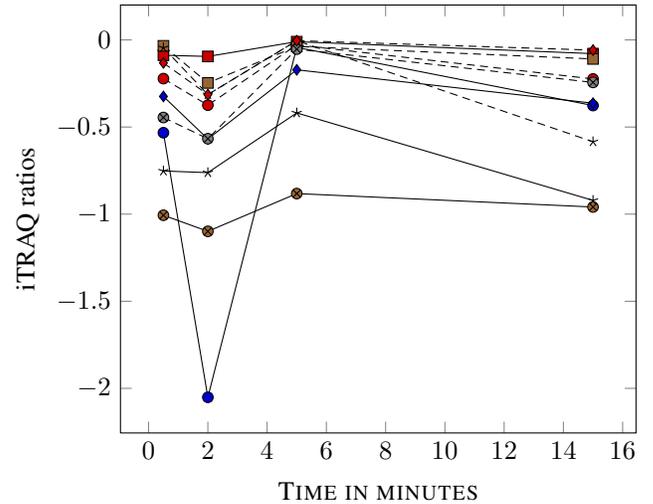

\begin{figure}[htb]
\begin{center}

\begin{tikzpicture}

\begin{axis}[legend pos=south east, xlabel=\textsc{Time in minutes},
ylabel=iTRAQ ratios
 ]

\addplot+[blue,sharp plot] coordinates
{(0.5,-.82347) (2,-.01537) (5,-.269) (15,-.30126)};

\addplot+[blue,sharp plot] coordinates
{(0.5,-.07214) (2,-.06779) (5,-.1223) (15,-.202)};

\addplot+[blue,sharp plot] coordinates
{(0.5,-.28757) (2,-.03476) (5,-.4205) (15,-.49238)};

\addplot+[blue,sharp plot] coordinates
{(0.5,-.107) (2,-.0528) (5,-.13271) (15,-.22938)};

\addplot+[blue,sharp plot] coordinates
{(0.5,-.23144) (2,-.09496) (5,-.5979) (15,-.6538)};

\addplot+[blue,sharp plot] coordinates
{(0.5,-2.32313) (2,-.01707) (5,-.84237) (15,-1.348)};

\addplot+[blue,sharp plot] coordinates
{(0.5,-.25482) (2,-.084116) (5,-.11353) (15,-.14131)};

\addplot+[blue,sharp plot] coordinates
{(0.5,-.16013) (2,-.07184) (5,-.11149) (15,-.11363)};

\addplot+[blue,sharp plot] coordinates
{(0.5,-.12436) (2,-.01438) (5,-.07323) (15,-.19777)};

\addplot+[blue,sharp plot] coordinates
{(0.5,-.26753) (2,-.00952) (5,-.08463) (15,-.2528)};

\addplot+[blue,sharp plot] coordinates
{(0.5,-.21983) (2,-.16667) (5,-.20643) (15,-.53814)};

\addplot+[blue,sharp plot] coordinates
{(0.5,-.04404) (2,-.02737) (5,-.15342) (15,-.41282)};

\addplot+[blue,sharp plot] coordinates
{(0.5,-.06831) (2,-.05738) (5,-.18911) (15,-.61936)};

\end{axis}
\end{tikzpicture}

\caption{\small \label{fig-cluster2} Example 2 of clustered data
points}

\end{center}
\end{figure}
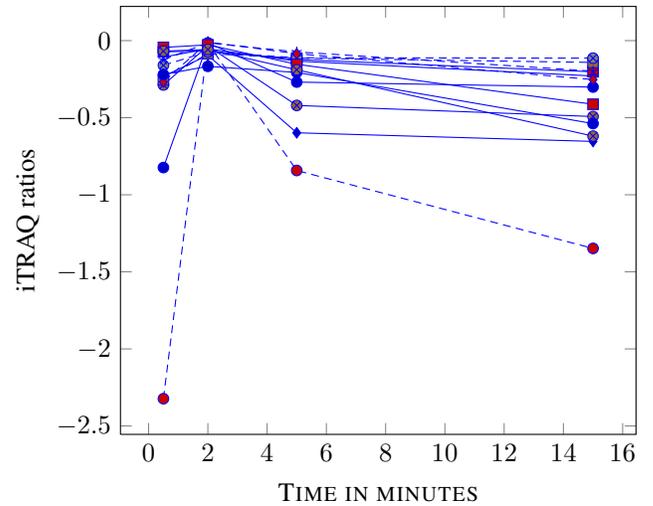

\begin{figure}[htb]
\begin{center}

\begin{tikzpicture}

\begin{axis}[legend pos=south east, xlabel=\textsc{Time in minutes},
ylabel=iTRAQ ratios
 ]

\addplot+[orange,sharp plot] coordinates
{(0.5,0.494945) (2,0.4003) (5,0.04289) (15,-.077638)};

\addplot+[orange,sharp plot] coordinates
{(0.5,0.1238) (2,0.086) (5,0.038) (15,-0.12303)};

\addplot+[orange,sharp plot] coordinates
{(0.5,0.181296572)(2,0.100554522)(5,-0.053492232)(15,-0.104892392)};

\addplot+[orange,sharp plot] coordinates
{(0.5,0.219031125)(2,0.048753247)(5,0.039241432)(15,-0.097937883)};

\addplot+[orange,sharp plot] coordinates
{(0.5,0.129818403)(2,0.098421283)(5,-0.193855926)(15,-0.282820011)};

\addplot+[orange,sharp plot] coordinates
{(0.5,0.194399549)(2,0.032494881)(5,-0.135315219)(15,-0.156403694)};

\addplot+[orange,sharp plot] coordinates
{(0.5,0.24227385)(2,0.12640148)(5,0.068863556)(15,-0.022068047) };

\addplot+[orange,sharp plot] coordinates
{(0.5,0.303375897)(2,0.195406301)(5,0.027186047)(15,-0.069271386) };

\addplot+[orange,sharp plot] coordinates
{(0.5,0.261044959)(2,0.064721309)(5,0.013088366)(15,-0.292908938)};

\addplot+[orange,sharp plot] coordinates
{(0.5,0.113037961)(2,0.101173442)(5,-0.051353691)(15,-0.280668299) };
\end{axis}
\end{tikzpicture}

\caption{\small \label{fig-cluster3} Example 3 of clustered data points}

\end{center}
\end{figure}
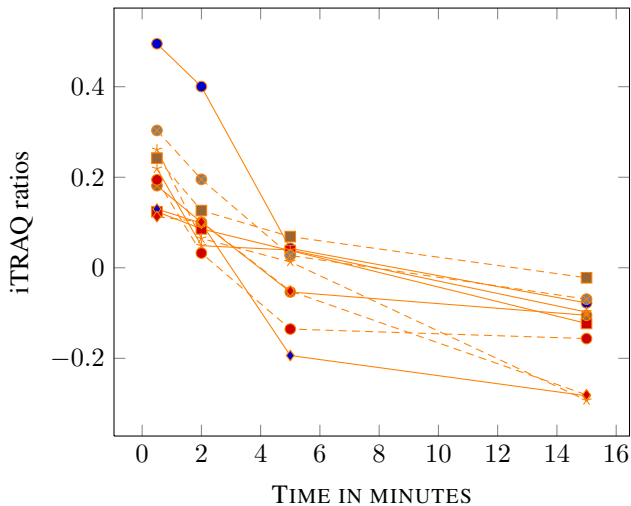

We were able to identify $16$ distinct clusters for
phosphopeptides with distinct temporal profiles.
We also performed subclustering of the data purely for biological analysis i.e. for quantification we wanted separate
clusters for the ratios that are negative all over the time
course, the ones that were positive and the ones that crossed from
positive to negative or negative or positive over the length of
the experiment. Some examples of the clusters are shown in Figs. \ref{fig-cluster1},
\ref{fig-cluster2} and \ref{fig-cluster3},each line presenting a peptide. As shown in Fig.
\ref{fig-cluster1}, even though the data points have high
variability in terms of distance of the points, the points have
been clustered accurately with respect to the pattern of the response. For all of the data points that were
clustered in Fig.\ref{fig-cluster1}, the ratios first decreased then increased and decreased
further in the last time point. Same can be observed for Fig.
\ref{fig-cluster2} and  \ref{fig-cluster3} that even though the
points have a lot of variability in terms of distance of the
points, they are clustered correctly in terms of patterns over the
time.

\section{Conclusion}\label{discussion}
We developed a new algorithm called \emph{TPM} for clustering time-course patterns
following step-inputs in biological systems for iTRAQ labeled
phosphopeptides. We tested the algorithm with data from
large-scale quantitative phosphoproteomics experiments for up to
$80,000$ data points and up to $100$ time point intervals.
Quantification used 8-plex iTRAQ and commercially available
software. These phosphopeptides were analyzed with our
algorithm in order to identify groups that changed in abundance with similar temporal responses
after vasopressin addition. The algorithm maps the data from a
Cartesian plane to a discrete binary plane and uses an efficient
dynamic programming technique to mine similar patterns after
mapping. The mapping allows clustering of similar time courses
that are temporally closer to each other. The algorithm identified
$16$ clusters of phosphopeptides with distinct temporal profiles in response to vasopressin. The
algorithm was also compared for quality to other standard
clustering techniques that have been used for similar experiments
in the literature. It was shown that the proposed algorithm
performed with significantly better accuracy (at least $99\%$ of clusters were
assigned correctly)  with the ability to handle large
data sets. These time-course clusters provide a starting point for
modeling of the signaling network. We believe that the
proposed algorithm will prove useful to the computational biology
and mass spectrometry community.

\subsection*{Acknowledgements}
Mass spectrometry was conducted in the National Heart, Lung and Blood Institute 
Proteomics Core Facility (director, Marjan Gucek). This work was funded by the 
operating budget of Division of Intramural Research, National Heart, Lung and 
Blood Institute, National Insitutes of Health (NIH), Project ZO1-HL001285.

\bibliography{mybib}
\bibliographystyle{ieeetr}
\end{document}